\documentstyle[12pt,equations,epsf]{article}

\begin{document}

\def\Journal#1#2#3#4{{#1} {\bf #2}, #3 (#4)}

\def\NCA{\em Nuovo Cimento}
\def\NIM{\em Nucl. Instrum. Methods}
\def\NIMA{{\em Nucl. Instrum. Methods} A}
\def\NPB{{\em Nucl. Phys.} B}
\def\PLB{{\em Phys. Lett.}  B}
\def\PRL{\em Phys. Rev. Lett.}
\def\PRD{{\em Phys. Rev.} D}
\def\ZPC{{\em Z. Phys.} C}

\def\beq{\begin{equation}}
\def\eeq{\end{equation}}
\def\bea{\begin{eqnarray}}
\def\eea{\end{eqnarray}}
\def\bit{\begin{itemize}}
\def\eit{\end{itemize}}
\def\wtil{\widetilde}
\def\what{\widehat}

\def\vev#1{\langle #1 \rangle}

\def\mev{~\mbox{MeV}}
\def\gev{~\mbox{GeV}}
\def\kev{~\mbox{keV}}
\def\tev{~\mbox{TeV}}
\def\eps{\epsilon}
\def\mw{m_W}
\def\mz{m_Z}
\def\leff{{\cal L}_{\rm eff}}
\def\mplanck{M_{\rm Pl}}
\def\nn{\nonumber}
\def\cntwo{\wt\chi_2^0}
\def\wt{\widetilde}
\def\gl{\wt g}
\def\mgl{m_{\gl}}
\def\lsim{\mathrel{\raise.3ex\hbox{$<$\kern-.75em\lower1ex\hbox{$\sim$}}}}
\def\gsim{\mathrel{\raise.3ex\hbox{$>$\kern-.75em\lower1ex\hbox{$\sim$}}}}
\def\fbi{~{\rm fb}^{-1}}
\def\etmiss{/ \hskip-6pt E_T \hskip6pt}
\def\cmone{\tilde{\chi}_1^-}
\def\mcpmone{m_{\cpmone}}
\def\cpone{\tilde{\chi}_1^+}
\def\cpmone{\tilde{\chi}_1^\pm}
\def\cnone{\tilde{\chi}_1^0}
\def\mcnone{m_{\cnone}}
\def\gev{~{\rm GeV}}
\def\mev{~{\rm MeV}}
\def\dmchi{\Delta m_{\tilde{\chi}_1}}
\def\epem{e^+e^-}
\def\anti{\overline}
\def\lam{\lambda}
\def\gam{\gamma}
\def\del{\delta}
\def\dbar{\overline D}
\def\dhat{\widehat D}
\def\htil{\widetilde h}
\def\phitil{\widetilde \phi}
\def\call{{\cal L}}
\def\lmix{\call_{\rm mix}}
\def\lkin{\call_{\rm kin}}
\def\lphi{{\call_\Phi}}
\def\half{{1\over 2}}
\def\quarter{{1\over 4}}
\def\vtot{V_{\rm tot}}
\def\vtotbar{\overline V_{\rm tot}}
\def\veff{V_{\rm eff}}
\def\vnorm{V_{\rm S}}
\def\sphys{s_{\rm phys}}
\def\msphys{m_{\sphys}}
\def\Phip{\Phi^\prime}
\def\fy{f_{\rm Y}}
\def\ldel{L}

\font\fortssbx=cmssbx10 scaled \magstep2
\hbox to \hsize{
$\vcenter{
\hbox{\fortssbx University of California - Davis}\medskip
}$
\vspace*{1.2cm}
$\vcenter{
\hbox{\bf UCD-00-10} 
\hbox{\bf hep-ph/0004058}
\hbox{April, 2000}
}$
}

\begin{center}
{\large\bf
Extra Dimension
Kaluza-Klein Excitations and Electroweak Symmetry Breaking}\footnote{To
appear in the Proceedings of the 7th International Symposium
on Particles, Strings and Cosmology, PASCOS 99.}
\rm
\vskip2pc
{\bf J.F. Gunion$^{a}$ and B. Grzadkowski$^b$}\\
\medskip
\small\it
$^a$ Davis Institute for High Energy Physics, 
University of California at Davis,\\
Davis, CA 95616, USA\\
$^b$ Inst. for Theoretical Physics, Warsaw University, Warsaw, Poland
\\
\end{center}
\vskip .5cm
\begin{abstract}
We review the possibilities that the Kaluza-Klein excitations of
graviton states induce electroweak symmetry breaking and
that electroweak
symmetry breaking could have a large impact on KK phenomenology.
\end{abstract}

In one popular approach~\cite{arkanietal} to extra dimensions
with length scales far larger than the inverse Planck mass~\cite{original},
ordinary particles are confined on a brane (having three spatial
dimensions) with gravity propagating in the bulk.
This talk\footnote{Presented by J. Gunion} summarizes
our paper~\cite{hgmix} 
showing that the KK modes could have a dramatic impact on 
electroweak symmetry breaking, possibly providing the EWSB
mechanism and how, in turn, EWSB could have a large impact on KK phenomenology.

We consider a theory with $\del$ extra dimensions of compactified size $\ldel$.
In the standard approach~\cite{hanetal,wellsetal},
a linearized approximation is employed in which
the metric tensor is expanded to first order in $\kappa=\sqrt{16\pi G_N}$ 
about the flat-space limit to obtain 
an effective Lagrangian and corresponding
equations of motion for the KK modes and Standard Model fields on the brane.
Focusing on the Higgs sector, there are three key components to the
resulting overall Lagrangian. 
(a) The Higgs Lagrangian, $\call_{\Phi}$, containing
an arbitrary Higgs potential $-V(\Phi)$ that for simplicity
we take to be a function of a single real Higgs scalar field $\Phi$.
The corresponding contribution to the energy
momentum tensor is $T_{\mu\nu}=\eta_{\mu\nu}V(\Phi)$.
(b) We also have the massive KK states, with masses
$m_{\vec n^2}={4\pi^2 \vec n^2\over \ldel ^2}$, 
where $\vec n=(n_1,n_2,\ldots,n_\del)$, and corresponding
mass-term Lagrangian, $\call_{\rm mass}^{\rm KK}$. (c) The third
key Lagrangian component is $\call_{\rm mix}$ which
contains the coupling between the graviton KK states
and the scalar field, which mixing 
is automatic since gravity couples to $T_{\mu\nu}$.

We find that the mixing term can lead to 
non-zero vacuum expectation values for the Higgs field
and the tensor and scalar KK fields in a certain class of models.
To see this, one considers the extremum of $\vtot=
V(\Phi)-\call_{\rm mass}^{\rm KK}-\lmix$.
After minimizing with respect to the KK fields and substituting into $\vtot$,
one finds a result denoted by $\vtotbar$:  $\vtotbar=V-\dbar V^2\,,$
where $V$ stands for $V(\Phi)$ and $\dbar=\kappa^2{\del-2\over\del+2}
\sum_{{\rm all}\,\vec n}{1\over m_n^2}$. The
extremum with respect to $\Phi$ is given by
${\partial \vtotbar\over\partial\Phi}
= {\partial V\over\partial\Phi}\left[1-2V{\dbar}\right]=0\,,$
with solutions $V={1\over 2\dbar}$ or
${\partial V\over\partial \Phi}=0$.
The 2nd solution corresponds to the usual minimum while the first
is of a very unusual nature, as we shall summarize. Whichever
extremum is appropriate, we denote the extremum values of $V$ and $\Phi$ by
$V_0$ and $v$, respectively. 
If we compare the values of $\vtotbar$ at the two extrema, we find
that the $V={1\over 2\dbar}$ extremum is preferred if 
$\dbar<0$. Writing~\cite{hanetal}
$\dbar={2\over M_S^4(\del-2)}$, where $M_S$ is an ultraviolet cutoff,
suggests that $\dbar>0$. However, the ultraviolet cutoff is
the point at which the physics of the string enters. The exact manner
in which the divergent sum is regularized is thus uncertain and either
sign for $\dbar$ is possible~\cite{hewett}. 
(A simple example is $\zeta$ regularization
for which $\dbar<0$ if $2-\del/2$ is odd.)
Thus, we consider $\dbar$ to simply be a parameter determined
by the detailed physics at the string scale.

Given a definite minimum, we must consider an appropriate quantum state 
expansion.  To this end, we first note that after substituting the 
vacuum expectation values for the KK fields into $\vtot$, one finds that
the derivative terms for $\Phi$ receive contributions
both from $\call_{\Phi}$ and from $\call_{\rm mix}$ and take the form
$
\call_\Phi+\lmix\ni \left(1-\dbar V_0\right)\half
\partial^\rho\Phi\partial_\rho\Phi,
$
so that, for the $V_0={1\over 2\dbar}$ minimum, half
of the usual
$\call^{\rm kin}_\Phi$ derivative term is canceled by $\lmix$ and
we must rescale $\Phi$ in order to have canonical
normalization for its kinetic energy term. We write
$
\Phi=\what\Phi\left(1-\dbar V_0\right)^{-1/2}\,.
$
The next step is to expand $\vtot= V(\Phi)-\call_{\rm mass}^{\rm KK}
-\lmix$ 
about the extremum. The quantum excitation for $\what\Phi$ is denoted
by $s$. The important KK states in the expansion of the spin-0 part
of the gravitational field are the non-zero `trace' components
with quantum excitations denoted $s_{\vec n}$. The excitations of the
tensor parts of the gravitational KK modes do not mix with these fields
and need not be considered.
The resulting form for $\vtot$ contains terms proportional
to $s_{\vec n}^2$,   $s^2$, and possibly (see below) $s_{\vec n}s$
(all $\vec n$). 

The physics at the two different extrema 
are quite different.  If ${\partial V\over\partial\hat\Phi}=0$,
the usual type of minimum, one finds
that there is no tree-level mixing
between the $s_{\vec n}$ and $s$. (This
is true even at the loop level.)  The Higgs and KK modes
remain in separate sectors. The case of $V_0={1\over 2\dbar}$ is more subtle.
The mass terms for the quantum fluctuations are of the form
$
\vtot\rightarrow{1\over 4}\sum_{\vec n}\left[m_{\vec n}^2 (s_{\vec n})^2+2\eps
s_{\vec n}\,s\right]\,,$
where $\eps\equiv 2\sqrt 2\kappa
\left[2(\del-1)\over 3(\del+2)\right]^{1/2} 
\left({\partial V\over\partial \what\Phi}\right)_{\what\Phi=\what v}\,.$
We then diagonalize $\vtot$. This is most easily done
for $\del\geq 4$, for which the off-diagonal entries in the
mass matrix are always
small compared to the diagonal entries. Even more importantly,
the upper cutoff in $m_{\vec n}^2$ to which we sum and
which dominates the relevant summations is 
much larger than $\eps$.
Thus, for the relevant matrix entries, the 
$s_{\vec n}$ KK states mix slightly with $s$ with a mixing angle 
$\theta_{\vec n}\sim -\eps {1 \over m_{\vec n}^2}$.
The physical eigenstate, $\sphys$, corresponding to the original $s$
ends up with mass given by
$\msphys^2\simeq
-{\eps^2\over 2}\sum_{\vec n} {1\over m_{\vec n}^2}
\sim -{8\over 3}\dbar {\del-1\over\del-2}
\left({\partial V\over\partial\what\Phi}\right)_{\what\Phi=\what v}^2\,$.
We note that the $\dbar<0$ requirement, needed to ensure
that we are expanding about a local minimum that is deeper
than the standard minimum, is also that which implies
a positive mass-squared for $s_{\rm phys}$.
In other words, mixing of $s$ with
the full tower of KK states and whatever physics is present
at the string scale to cutoff the ultraviolet divergence of 
$\sum_{\vec n}{1\over m_{\vec n}^2}$ is critical to obtaining $\msphys^2>0$.

It is very significant that the $\dbar<0$ minimum
can yield nonzero $v$ at the minimum even if $V(\Phi)$ itself does not
have a minimum with $\Phi\neq0$.
In particular, $V(\Phi)=\half m^2 \Phi^2+\Xi$
is entirely satisfactory,
provided $V_0=\half m^2v^2+\Xi={1\over 2\dbar}<0$, where 
$v=\sqrt 2\hat v$.
For this form of $V(\Phi)$ we have 
$
\msphys^2=-{32\over 3}\dbar {\del-1\over\del-2}m^4\what v^2\,.$
For natural choices of $m$, 
to achieve $V_0={1\over 2\dbar}$ it is necessary that $\Xi$
be negative with absolute magnitude of order $M_S^4$, a seemingly
natural possibility.

Let us now turn to vector boson mass generation.
There are two contributions; one coming from $\lmix$ and
the other from the standard $\call^{\rm kin}_\Phi$ kinetic energy
portion of the scalar Lagrangian. After substituting the vacuum expectation
values for the KK fields into $\call_{\rm mix}$ and rescaling the $\Phi$
fields as described earlier, one finds 
\beq
\lmix+\call^{\rm kin}_\Phi\ni
{1\over4}g^2(\what v+s)^2W_\rho^+W^{-\,\rho}
+{1\over 8}\left(g^2+g^{\prime\,2}\right)(\what v+s)^2Z_\rho Z^\rho\,.
\label{wzlmass}
\eeq
At this point, it is important to note 
that the $W$ and $Z$ fields themselves do not need to be rescaled
in order to have canonical normalization. This is because, for any given 
$V_0$, $\call_{\rm mix}={1\over 2}\dbar V_0T_\mu^\mu$ and
the $F^{\rho\sigma}F_{\rho\sigma}$
kinetic energy part of the vector field contribution
to the full $T_\mu^\mu$ is zero.
Thus, we can proceed to read off masses from Eq.~(\ref{wzlmass}),
from which it is obvious that one obtains the usual results for
the $W$ and $Z$ masses. As for the couplings
of $WW$ and $ZZ$ to $\sphys$, Eq.~(\ref{wzlmass}) shows that $s$
has the standard couplings. For $\dbar<0$, we rotate to $\sphys$, and find
a small correction to this coupling of order 
$gm_W(\what v^2m^2/M_S^4)$. This, and similar small corrections
due to the small rotation from $s$ to $\sphys$, will be neglected
in the following discussion.

The fermion sector yields a surprise. First, examination of the kinetic energy
terms shows that we must rescale to
$
\what\psi=\left(1-{3\over 2}\dbar V_0\right)^{1/2}\psi\,.
$
As regards the Yukawa coupling, for which we use the notation
$\call_\psi\ni -\fy \overline\psi\psi\Phi$, after rescaling
both $\Phi$ and $\psi$ one finds
$
\call_\psi+\lmix\ni 
-\left(1-2\dbar V_0\right) \left(1-{3\over 2}\dbar V_0\right)^{-1}
\left(1-\dbar V_0\right)^{-1/2}\fy \overline{\what\psi}\what\psi\what\Phi \,,
$
where the $(1-2\dbar V_0)$ arises even before
rescaling as a result of combining $\call_\psi$ and $\lmix$.
If $1-2\dbar V_0\neq 0$, then this simply amounts to a redefinition
of the Yukawa coupling strength $\fy $, which does 
not affect the standard relation between
the $s \what\psi\anti{\what \psi}$ 
coupling and the mass $m_{\what\psi}$ induced
by $\what v$. 
However, for the $\dbar<0$ minimum,
$1-2\dbar V_0=0$ and it appears that the Yukawa interaction is automatically
zeroed (at tree level).  

We next wish to point out that the basic gauge-theory interaction
strengths are not altered by the rescaling required when $V_0\neq 0$.
For example, consider the interaction of the fermionic $\psi$
field with a vector field.
After the rescaling, $\call_\psi+\call_{\rm mix}\ni  
\overline{\what\psi} i\gamma^\rho D_\rho \what\psi$,
where $D_\rho$ is the usual gauge covariant derivative, 
is of canonical form.  This,
in combination with the fact that there is no rescaling for the vector fields
contained in $D_\rho$ implies that the $W\what\psi\anti{\hat \psi}$ and
$Z\what\psi\anti{\what\psi}$ couplings are the same as always. 
The same remarks apply also to the interactions of the Higgs fields
with the vector fields.  Indeed, after rescaling,
the Higgs kinetic energy terms have a canonical
normalization; by making
the derivatives covariant $\call_\Phi+\call_{\rm mix}\ni
\sum \half (D_\rho\what\Phi)(D^\rho\what\Phi)$, which 
(when we expand $\what\Phi=\what v+s$) 
leads to Eq.~(\ref{wzlmass}) and, thence, standard
gauge couplings for $s$. Clearly,
the gauge structure of the theory is being preserved precisely because
the vector fields do not require rescaling.

If $\dbar>0$, then electroweak symmetry
breaking is only possible if $V$ itself has a minimum for non-zero
$\Phi$. A typical form is $V=\lambda(\Phi^2-v^2)^2+\Xi$,
leading to $V_0=\Xi$. In this case, it
is important to realize that the Higgs self interactions induced
by such a potential will not be related in the usual way to
the Higgs mass if $V_0\neq 0$ and the Higgs fields are rescaled.

An important question is how to resolve $V_0\neq 0$
(required to be of order $M_S^4$ in magnitude if $\dbar<0$
and most naturally of this magnitude even if $\dbar>0$) 
with the known fact that the  
vacuum energy on the brane (i.e in our world of three
spatial dimensions) is very small.
We first note that adding an explicit cosmological constant 
that exists only on the brane  is simply equivalent to shifting
the value of the constant $\Xi$ as already included in the Higgs potential.
Possible solutions to the problem thus require introducing a nearby brane
or other source of bulk cosmological constant that cancels $V_0$
on our brane.    
For the natural magnitude of $|V_0|\sim M_S^4$ (as certainly
required for $\dbar<0$), the cancellation between $V_0$ and the cosmological
constant coming from the bulk
must be essentially exact.  However, we do not regard this as being
unreasonable given that all these quantities will be determined
by the ultimate string theory which might well have such an exact 
cancellation built in by means of symmetry or dynamics.
An unresolved question
is whether the bulk physics can be introduced 
in such a way as to have minimal impact
on the linearized quantum gravity mode expansion employed in our treatment.

There are some very important phenomenological implications of our results.
First, consider the popular KK phenomenology.
We have argued that $V_0\neq 0$ is possible even if $\dbar>0$,
and obviously it is required if $\dbar<0$. If $V_0\neq0$,
then the KK phenomenology given in the literature will be altered.  
The crucial point is 
that if $V_0\neq 0$ then the Higgs scalar fields
and the fermionic fields must be rescaled to achieve canonical
normalization. In addition, in the $V_0={1\over 2\dbar}$ minimum
with $\partial V/\partial\Phi\neq0$ there will be Higgs-KK mixing.
Both effects will modify the KK couplings to the physical 
states.\footnote{However, the KK-mode induced mixing of $s_{\vec n}$
with $s$, present at tree-level if $\dbar<0$, does not lead to 
experimentally significant modifications because of the small 
size of the mixing 
angles $\theta_{\vec n}\sim -{\epsilon\over m_{\vec n}^2}$ given that
$\eps\propto \kappa$. It is
only if one performs experiments at energies of order the cutoff
scale $M_S$ that the cumulative effects of these small mixings might
become significant.}
We have already noted that the vector
fields are not rescaled, implying that vector-vector-KK interactions
are not altered if $V_0\neq 0$. In the fermion case,
the Feynman rules~\cite{hanetal} for fermion-antifermion-KK interactions 
will be modified by a factor of
$(1-{3\over 2}\dbar V_0)^{-1}$. For the $V_0={1\over 2\dbar}$ minimum,
the coupling strengths of the KK modes to $\what\psi
\overline{\what\psi}$ are then obtained by multiplying
the Feynman rules by a factor of 4.
In the Higgs field case, the Feynman rules~\cite{hanetal}
for KK mode coupling to two Higgs fields must be multiplied
by $(1-\dbar V_0)^{-1}$, which is a factor of 2 for the $V_0={1\over 2\dbar}$
minimum. A sampling of the consequences are the following.
(A)
The effective contact interaction generated by virtual KK exchanges 
is multiplied
by a factor of 16 for 4-fermion interactions and by a factor
of 4 in the case of vector-vector-fermion-fermion interactions.
This means that the experimental constraints on $M_S$ will
be increased by a factor of 2 ($\sqrt 2$) in the respective cases.
(B)
The amplitude for radiating a KK excitation from a fermion (vector boson)
is increased by a factor of 4 (2). This means that the upper bound on
$M_S$  extracted from experimental limits on KK radiation
must be re-evaluated.  
 
The modifications to Higgs phenomenology are also dramatic in
the case of the $\dbar<0$ minimum.  
Although the couplings of the physical Higgs boson, $\sphys$,
to vector bosons are essentially
the same as in the Standard Model, $\sphys$ has no fermionic couplings
at tree level. Its primary Standard Model decay modes will
then be to $WW^{(*)}$ and $ZZ^{(*)}$ at higher masses ($\msphys>m_W$ or so)
with $\sphys\to\gam\gam$ decays being very important at low mass. Further,
(invisible) decays into two KK $s_{\vec n}$ excitations can be substantial
(perhaps even dominant). 

To summarize, we have found that mixing between the Higgs
sector and the KK modes could provide a 
source for electroweak symmetry breaking
even in the absence of tree-level Higgs self interactions. The proposed
mechanism arises automatically if the KK mode sum 
$\sum_{\vec n}{\kappa^2\over m_{\vec n}^2}\propto \dbar$ is cutoff
at the string scale in such a way that $\dbar<0$. Indeed,
electroweak symmetry breaking occurs when $\dbar<0$ whatever
the value of the actual string cutoff scale, $M_S\sim |\dbar|^{-1/4}$,
so long as $M_S$ is sufficiently below $\mplanck$ that the
effective theory we employ can be defined.
Even if the KK modes are not responsible for electroweak symmetry breaking,
the phenomenology of the contact interactions and missing
energy processes which they mediate 
could be greatly modified if the Higgs potential
vacuum expectation value is of order $M_S^4$, as is entirely possible
and perhaps even natural in the string theory context. Finally,
Higgs phenomenology is very substantially modified in the
case of the $\dbar<0$ minimum.

\bigskip  

\centerline{\bf Acknowledgements}
\bigskip
This work was supported in part by the U.S. Department of Energy,
the U.C. Davis Institute for High Energy Physics, the State Committee for
Scientific Research (Poland) grant No. 2~P03B~014~14 and the Maria
Sklodowska-Curie Joint Fund II (Poland-USA) grant No. MEN/NSF-96-252.

\noindent

\end{document}